\documentclass[12pt,a4]{article}
\usepackage{amsthm,amsmath,latexsym,amssymb,amsfonts,amscd}
\usepackage{graphics,lscape,fancyhdr,array,stmaryrd,euscript}
\usepackage{empheq}
\usepackage{sidecap}
\usepackage{ifpdf}
\usepackage{verbatim}
\usepackage{color}
\usepackage{relsize}
\usepackage{tikz-cd}
\numberwithin{equation}{section}
\usepackage{hyperref,setspace}
\usepackage{soul}
\usepackage[numbers,sort&compress]{natbib}
\usepackage[nottoc]{tocbibind}
\usepackage{skak}

\newcommand{\kb}{\bar{k}}

\newcommand{\be}{\begin{equation}}
\newcommand{\ee}{\end{equation}}
\newcommand{\lvec}{\boldsymbol{\ell}}

\newcommand{\besubeqs}{\begin{subequations}}
\newcommand{\esubeqs}{\end{subequations}}

\newcommand{\pvec}{{\boldsymbol{p}}}
\newcommand{\qvec}{{\boldsymbol{q}}}
\newcommand{\kvec}{{\boldsymbol{k}}}
\newcommand{\pb}{{\bar{p}}}

\newcommand{\tr}{{\text{Tr}}}

\newcommand{\PP}{{\mathbb{P}}}

\newcommand{\PPb}{{\overline{\mathbb{P}}}}

\begin{document}

\pagenumbering{gobble}
\hfill
\vskip 0.01\textheight
\begin{center}

{\Large\bfseries 
One-loop Finiteness of 
Chiral Higher Spin Gravity}\\

\vskip 0.03\textheight

Evgeny Skvortsov,${}^{\symking,\dagger}$ Tung Tran,${}^{\symbishop,\symking}$

\vskip 0.04\textheight

{\em$^{\symking}$  Albert Einstein Institute, \\
Am M\"{u}hlenberg 1, D-14476, Potsdam-Golm, Germany}

\vspace{5pt}
{\em ${}^{\symbishop}$ Arnold Sommerfeld Center for Theoretical Physics\\
Ludwig-Maximilians University Munich\\
Theresienstr. 37, D-80333 Munich, Germany}\\

\vspace{5pt}
{\em$^{\dagger}$ Lebedev Institute of Physics, \\
Leninsky ave. 53, 119991 Moscow, Russia}\\

\end{center}

\vskip 0.02\textheight

\begin{abstract} 
One of the main ideas behind Higher Spin Gravities is that the higher spin symmetry is expected to leave no room for counterterms, thereby eliminating UV divergences that make the pure gravity non-renormalizable. However, until recently it has not been clear if such a mechanism is realized. We show that Chiral Higher Spin Gravity is one-loop finite, the crucial point being that all one-loop $S$-matrix elements are UV-convergent despite the fact that the theory is naively not renormalizable by power counting. For any number of legs the one-loop $S$-matrix elements coincide with all-plus helicity one-loop amplitudes in pure QCD and SDYM, modulo a certain higher spin dressing, which is an unusual relation between the non-gravitational theories and a higher spin gravity.
\end{abstract}

\newpage
\section*{Introduction}
\pagenumbering{arabic}
\setcounter{page}{2}
It is fair to say that there is no widely accepted solution to the Quantum Gravity Problem that has all the desired features. Therefore, pursuing different approaches to the problem and constructing various simple toy models should help to shed more light and chart the landscape of consistent quantum gravity models. One promising approach to get simple models of this kind is via Higher Spin Gravity (HSG), i.e. by trying to extend the graviton with massless higher spin fields. In the paper we prove that Chiral HSG \cite{Metsaev:1991mt,Metsaev:1991nb,Ponomarev:2016lrm,Skvortsov:2018jea,Skvortsov:2020wtf} is one-loop finite. A remarkable fact is that the multitude of non-renormalizable interactions, including the two-derivative graviton self-interaction, conspire to give UV-convergent integrals thanks to the higher spin symmetry \cite{Skvortsov:2018jea,Skvortsov:2020wtf}. The final result for the one-loop amplitudes coincides, modulo certain higher spin factors, with the all-plus helicity amplitude in QCD and SDYM.  

The idea behind HSG is that a large gauge symmetry associated with a higher spin extension of general relativity will leave no room for relevant counterterms, \cite{Fronsdal:1978rb}. Another idea behind HSG is to overcome the restriction $\mathcal{N}\leq8$ on the number of supersymmetries in supergravities \cite{Fradkin:1986ka}. While going beyond $\mathcal{N}=8$ requires higher spin fields one may also embed the graviton into a multiplet of only bosonic higher spin fields \cite{Flato:1978qz}. The higher spin symmetry should be powerful enough to guarantee the consistency at the quantum level even without supersymmetry. One key property of HSG in dimension four and greater is that the minimal multiplet that contains the graviton and at least one higher spin field is always infinite \cite{Flato:1978qz,Berends:1984wp,Fradkin:1986ka,Maldacena:2011jn,Boulanger:2013zza}. Therefore, HSG are not quite conventional field theories, being somewhat close to string theory where massive higher spin states play an important role.

It has been difficult to construct HSG’s even as classical theories due to numerous inconsistencies usually caused by the presence of massless higher spin particles. The issues arise both in flat  \cite{Weinberg:1964ew,Coleman:1967ad,Fotopoulos:2010ay,Bekaert:2010hp,Roiban:2017iqg} and, as was shown recently, anti-de Sitter spaces \cite{Dempster:2012vw,Bekaert:2015tva,Maldacena:2015iua,Sleight:2017pcz,Ponomarev:2017qab}. This shows that the main challenges of HSG have little to do with the cosmological constant:\footnote{One can, however, find certain technical distinctions between the higher spin problem in flat space and in (anti)-de Sitter one. For example, the gravitational interactions of higher spin fields do exist both in $AdS_4$ \cite{Fradkin:1986qy} and, surprisingly, in flat space within the $4d$ light-cone approach, but there does not seem to exist their covariant formulation in terms of Fronsdal fields \cite{Conde:2016izb}. See e.g. \cite{Bekaert:2010hw,Conde:2016izb} for an account of discrepancies. } massless higher spin fields require higher derivative interactions \cite{Bengtsson:1983pd,Bengtsson:1983pg,Berends:1984wp}, while higher spin symmetry can mix both spins and derivatives. The latter being in conflict with the very basic principles of quantum field theory and is insensitive to the value of the cosmological constant. Therefore, HSGs are very scarce. An optimistic view on these issues is that HSGs can be good probes of the Quantum Gravity Problem since the quantum issues are, to some extent, pushed to the classical level. Indeed, if the higher spin symmetry forbids any relevant counterterms, it would suffice to have a classical HSG theory and an appropriate regularization to give a consistent quantum gravity model.  

A list of HSG that are well-defined and have actions is very short: topological massless theories in $3d$ \cite{Blencowe:1988gj,Bergshoeff:1989ns,Campoleoni:2010zq,Henneaux:2010xg}; higher spin extensions of conformal gravity in $3d$ \cite{Pope:1989vj,Fradkin:1989xt,Grigoriev:2019xmp} and $4d$ \cite{Segal:2002gd,Tseytlin:2002gz,Bekaert:2010ky}; Chiral HSG \cite{Metsaev:1991mt,Metsaev:1991nb,Ponomarev:2016lrm}. The latter is, therefore, the only HSG with propagating massless higher spin fields where direct quantum checks are possible.\footnote{Holographic HSGs face certain difficulties \cite{Bekaert:2015tva,Sleight:2017pcz,Ponomarev:2017qab}, but an (nonstandard) action can be reconstructed from the CFT dual \cite{Das:2003vw,deMelloKoch:2018ivk}. Another interesting proposal is \cite{Sperling:2017dts}. We are not aware of any other proposal that would qualify, i.e. to give systematic well-defined predictions for interactions. } Therefore, we focus on Chiral Theory that has a rather simple action, although in the light-cone gauge. Moreover, we investigate  Chiral Theory in flat space for simplicity reason since the UV-properties should not depend much on the curvature. Note that the conformal HSGs have both flat and anti-de Sitter spaces as consistent backgrounds and Chiral HSG has both flat and anti-de Sitter versions \cite{Metsaev:2018xip,Skvortsov:2018uru}. This again indicates that the cosmological constant plays no role.

The action of Chiral Theory \cite{Metsaev:1991mt,Metsaev:1991nb,Ponomarev:2016lrm,Skvortsov:2018jea,Skvortsov:2018uru} is known in the light-cone gauge. Following the same logic as string theory in the light-cone gauge \cite{Goddard:1973qh} one attempts to construct the generators of the Poincare algebra \cite{Bengtsson:1983pd,Bengtsson:1983pg}, where the nontrivial relations to check are
\begin{align}\tag{$\ast$}
[J^{a-},J^{c-}]&=0\,, &
[J^{a-},P^{-}]&=0\,.\label{stringeq}
\end{align}
The action is then obtained from the Hamiltonian $P^-$. Eq. \eqref{stringeq} fixes both the spectrum of the theory and the interactions in $P^-$ and $J^{a-}$. One important property of Chiral Theory is its minimality: it is the smallest higher spin theory that incorporates at least one higher spin field with a nontrivial self-interaction and the graviton. In addition, Chiral Theory is not an isolated one. It has to be a closed subsector of any other higher spin theory in four dimensions. Technically, if we had an action of a theory that is dual to the large-$N$ free or critical vector model \cite{Klebanov:2002ja}, or more generally to Chern-Simons Matter theories \cite{Giombi:2011kc}, we could erase most of the terms and get the Chiral Theory's action.\footnote{It would be interesting to extract it also from \cite{Das:2003vw,deMelloKoch:2018ivk}, which are close to the light-cone formulation.} This remarkable property is special to four dimensions and is reminiscent of the relation between Yang-Mills Theory and its self-dual truncation --- in the light cone gauge the latter is obtained by erasing the quartic and half of the cubic vertices \cite{Chalmers:1996rq}. The analogy can be made more precise due to the hidden self-duality of Chiral Theory \cite{Ponomarev:2017nrr}. Another closely related example is self-dual gravity \cite{Siegel:1992wd,Krasnov:2016emc}.

Our main result is one-loop finiteness of Chiral Theory. A somewhat naive argument is to use unitarity cuts. It was shown in \cite{Skvortsov:2018jea,Skvortsov:2020wtf} that the physical (on-shell) tree-level amplitudes do vanish in Chiral Theory. This vanishing is a result of a highly nontrivial cancellation among all Feynman diagrams. Therefore, all one-loop cuts should vanish and the one-loop amplitudes have to be finite rational expressions, as in self-dual Yang-Mills or for all-like helicity in QCD \cite{Bern:1993sx,Bern:1993qk,Mahlon:1993fe}. However, the Hamiltonian is not Hermitian as in SDYM and some subtle higher spin features can make the low-spin inspired arguments invalid. Therefore, we directly approach the one-loop $S$-matrix elements for any number of legs. Nevertheless, a modified unitarity argument allows us to represent the sum over all relevant one-loop Feynman graphs as minus the sum over all possible insertions of the self-energy correction into the tree diagrams. The latter are UV-finite and do not vanish. Therefore, the total one-loop integrand does not lead to any UV-divergences. 

The final result is that the complete $n$-point one-loop $S$-matrix element consists of three factors: the all-like helicity one-loop amplitude in QCD (or self-dual Yang-Mills), which can be anticipated from \cite{Ponomarev:2017nrr};\footnote{As a side remark, the computation in the paper, after erasing the higher spin modes, can give a simple way to compute one-loop amplitudes in self-dual Yang-Mills. } a certain higher spin dressing --- an overall kinematical factor that accounts for the helicities on the external legs; a purely numerical factor of the total number of degrees of freedom:
\begin{align}\tag{$\Diamond$}\label{result}
    \Gamma_{\text{Chiral HSG, 1-loop}}&= \Gamma^{++...+}_{\text{QCD, 1-loop}} \times \left[\,\parbox{3.5cm}{kinematical \\ higher spin dressing}\right] \times \sum_{\lambda=-\infty}^{+\infty} 1\,.
\end{align}
Direct evaluation of one-loop integrals with $2$, $3$ and $4$ legs reveals the nuts and bolts of how higher spin fields eliminate UV-divergences: the specific structure of higher derivative interactions helps to factor enough momenta out of the integrand to make the integral UV-convergent, which is somewhat reminiscent of $\mathcal{N}=4$ Yang-Mills Theory \cite{Mandelstam:1982cb,Brink:1982wv} where one power of the momentum suffice.
The total number of effective degrees of freedom $\sum_{\lambda=-\infty}^{+\infty} 1$ should be regularized to $0$ according to \cite{Beccaria:2015vaa}. The final one-loop scattering amplitude vanishes in Chiral Theory, which is consistent with the Weinberg and Coleman-Mandula theorems. We note that the tree-level holographic $S$-matrix of Chiral Theory in $AdS_4$ does not vanish and is related \cite{Skvortsov:2018uru} to the correlation functions in Chern-Simons Matter Theories, which supports the dualities they were conjectured to exhibit \cite{Giombi:2011kc, Maldacena:2012sf, Aharony:2012nh,Aharony:2015mjs,Karch:2016sxi,Seiberg:2016gmd}.  

The paper is organized in a simple way. In section \ref{sec:chiral} we briefly describe Chiral Theory. In section \ref{sec:oneloop} the one-loop corrections are shown to be UV-finite and are then computed. Conclusions and discussion can be found in section \ref{sec:conclusions}.

\section{Chiral Higher Spin Gravity}
\label{sec:chiral}
We refer to \cite{Metsaev:1991mt,Metsaev:1991nb,Ponomarev:2016lrm,Ponomarev:2016jqk,Ponomarev:2017nrr,Skvortsov:2020wtf} for the detailed description of Chiral Theory and e.g. to \cite{Metsaev:2005ar} for the systematical introduction to the light-cone approach. 

A massless spin-$s$ field in $4d$ can be described by two complex conjugate scalars $\Phi^{+s}(\pvec)$ and $\Phi^{-s}(\pvec)$, which represent the helicity eigen states. Throughout the paper we work in momentum space and $\pvec=(p^+,p^-,p,\bar{p})$,  $\pvec^2=2p^+p^-+2p\pb$. It is convenient to use $\beta\equiv p^+$ instead of $p^+$. The light-cone notation might look cumbersome, but there is a direct link to the spinor-helicity formalism \cite{Chalmers:1998jb,Chakrabarti:2005ny,Chakrabarti:2006mb,Ananth:2012un,Bengtsson:2016jfk,Ponomarev:2016cwi}. Given several momenta $\pvec_k$, one introduces $\PPb_{km}=\pb_k\beta_m-\pb_m\beta_k$ and similarly for $\PP_{km}$. The two-component spinors are defined as
\begin{align}
|i] &= 2^{1/4}\left(\begin{array}{c}
  \pb_i \beta_i^{-1/2} \\
 - \beta^{1/2}_i
\end{array}\right)\,, &
|i\rangle &= 2^{1/4}\left(\begin{array}{c}
  p_i \beta_i^{-1/2} \\
 - \beta^{1/2}_i
\end{array}\right)\,.
\end{align}
The contractions can be expressed as
\begin{equation}
\label{9a1}
[i  j] = \sqrt{\frac{2}{\beta_i\beta_j}}\PPb_{ij}\,, \qquad\quad \langle ij \rangle = \sqrt{\frac{2}{\beta_i\beta_j}}\PP_{ij}\,,
\end{equation}
As is shown in \cite{Metsaev:1991mt,Metsaev:1991nb}, given three massless states with helicities $\lambda_{1,2,3}$, there exists a unique cubic vertex provided $\lambda_1+\lambda_2+\lambda_3>0$
\begin{align}
\frac{\PPb^{\lambda_1+\lambda_2-\lambda_3}_{12}\PPb^{\lambda_1+\lambda_3-\lambda_2}_{13}\PPb^{\lambda_2+\lambda_3-\lambda_1}_{23}}{\beta_1^{\lambda_1}\beta_2^{\lambda_2}\beta_3^{\lambda_3}} \sim\frac{\PPb^{\lambda_1+\lambda_2+\lambda_3}}{\beta_1^{\lambda_1}\beta_2^{\lambda_2}\beta_3^{\lambda_3}}\sim 
    [12]^{\lambda_1+\lambda_2-\lambda_3}[23]^{\lambda_2+\lambda_3-\lambda_1}[13]^{\lambda_1+\lambda_3-\lambda_2}\,,
\end{align}
where modulo momentum conservation all $\PPb_{12}$, $\PPb_{23}$, $\PPb_{31}$ are equivalent to
\begin{align} \label{PP1}
    \PPb=\frac13\left[ (\beta_1-\beta_2)\bar{p}_3+(\beta_2-\beta_3)\bar{p}_1+(\beta_3-\beta_1)\bar{p}_2\right]\,.
\end{align}
The light-cone vertex, which is off-shell, leads to the canonical spinor-helicity expression for an amplitude \cite{Benincasa:2007xk,Benincasa:2011pg}. If $\lambda_1+\lambda_2+\lambda_3=0$ then only the scalar cubic vertex is allowed, i.e. $\lambda_{1,2,3}=0$. Note that these constraints follow from locality, i.e. neither $P^-$ nor $J^{a-}$ are allowed to have inverse powers of the transverse momenta $p$, $\pb$. Analogously, one can write down the vertices for $\lambda_1+\lambda_2+\lambda_3<0$ that are expressed in terms of $\PP$ and $\langle ij\rangle$. 

The spectrum of Chiral Theory contains all integer or at least all even spins. Yang-Mills gaugings are also possible \cite{Metsaev:1991nb,Skvortsov:2020wtf} with the pattern that is reminiscent of the Chan-Paton method in string theory, see also \cite{Konstein:1989ij}. To this end, one makes $\Phi^\lambda$ to take values in the matrix algebra, the matrix indices being implicit. The action has the following simple form
\begin{align}
\begin{aligned}\label{eq:chiralaction}
S=-\sum_{\lambda\geq0}\int (\pvec^2)\mathrm{Tr}[\Phi^{\lambda}(\pvec)^\dag \Phi^\lambda(\pvec)] +\sum_{\lambda_{1,2,3}}\int C_{\lambda_1,\lambda_2,\lambda_3} V(\pvec_1,\lambda_1;\pvec_2,\lambda_2;\pvec_3,\lambda_3)\,,
\end{aligned}
\end{align}
where $\mathrm{Tr}$ is the trace over the optional matrix indices and the vertices are the standard ones
\small
\begin{equation}\label{eq:generalvertex}
    V(\pvec_1,\lambda_1;\pvec_2,\lambda_2;\pvec_3,\lambda_3)=\frac{\PPb^{\lambda_1+\lambda_2+\lambda_3}}{\beta_1^{\lambda_1}\beta_2^{\lambda_2}\beta_3^{\lambda_3}}\tr[\Phi^{\lambda_1}_{\pvec_1}\Phi^{\lambda_2}_{\pvec_2}\Phi^{\lambda_3}_{\pvec_3}]\delta^4(\pvec_1+\pvec_2+\pvec_3)\,.
\end{equation}
\normalsize
There coupling constants are fixed in a unique way to be
\begin{equation}\label{eq:magicalcoupling}
    C_{\lambda_1,\lambda_2,\lambda_3}=\frac{\kappa\,(l_p)^{\lambda_1+\lambda_2+\lambda_3-1}}{\Gamma(\lambda_1+\lambda_2+\lambda_3)}\,,
\end{equation}
where $\kappa$ is dimensionless and plays no role in the paper and $l_p$ has the dimension of length. The action contains only the chiral half of each vertex and is missing the conjugate vertices built out of $\PP$. The lower spin terms include: $C^{+1,+1,-1}$ corresponds to the chiral half of the Yang-Mills cubic vertex $FAA$; $C^{+2,+2,-2}\sim l_p$ corresponds to the chiral half of the two-derivative graviton cubic vertex extracted from $\sqrt{g}R$.  Certain higher derivative 'counterterms' for low spin fields are also present: $C^{+1,+1,+1}$ is the cubic three-derivative term built out of the self-dual component $F^+$ of the Yang-Mills field strength $F_{\mu\nu}$, $\mathrm{Tr}[F^+F^+F^+]$ and we omit the Lorentz indices; $C^{+2,+2,+2}$ is the chiral half of the Goroff-Sagnotti \cite{Goroff:1985th} counterterm 
\begin{equation}\label{Div}
\int \sqrt{g}\, R_{\mu\nu\rho\sigma}
R^{\rho\sigma\lambda\tau} R_{\lambda\tau}^{~~~\mu\nu}\,,
\end{equation}
There are also plenty of interactions involving higher spin fields. It is also important that the scalar field is a part of the spectrum, but its cubic self-coupling is absent. In some sense the action contains all possible reasonable interactions and all of them play both the roles of basic interactions and of the necessary counterterms at the same time. This is consistent with the general expectation that all possible terms should be present in the action. 

For simplicity we will work with the large-$N$ limit of the $U(N)$ Chiral Theory. In the latter case, fields $\Phi^\lambda$ are taken to be $N\times N$ matrices, the spin-one states $\Phi^{\pm 1}$ turn into a $U(N)$ Yang-Mills field and all other components of the higher spin multiplet become charged with respect to it. The formulas below are also valid, up to self-evident $N$-factors and permutations, for the version of the theory that has all integer spins.

\section{One-loop Finiteness}
\label{sec:oneloop}
The main argument is a generalization of the one used in \cite{Chakrabarti:2005ny} to compute $\Gamma^{++++}$ amplitude in pure QCD at one loop. We will prove that the sum of all one-loop integrands for the $n$-particle one-loop $S$-matrix element is equal to the sum over the self-energy insertions into various lines of the tree-level diagrams. The latter is UV-finite and we will compute it.  

Let us take a sum of integrands of all one-loop Feynman diagrams with $n$ external on-shell momenta $\pvec_i$, $\pvec_i^2=0$. We denote this sum $F$. The loop momentum is $\lvec$. $F$ is a rational function of momenta $\pvec_i$, $\lvec$. Note, that the vertices do not contain the minus-component of the momenta. Therefore, $p^-_i$, $\ell^-$ appear only in the denominators, as a part of the propagator, $\pvec^2=2p^+p^- +2 p\pb$. Now, $F$, as a function of $\ell^-$, vanishes at infinity and has only simple poles. The poles correspond to some momenta along the loop going on-shell in various diagrams that contribute to $F$. Since the loop momenta is to be integrated over, there is a an ambiguity in the momenta assigned to the lines going around any loop. Indeed, we can simply add any amount $\qvec$ to all momenta of the loop. We would like to choose the momenta around the loop in such a way that the residues of $F$ at the poles in $\ell^-$ give the complete $(n+2)$-point tree-level amplitude:
\begin{align*}
    F=\sum\parbox{3.cm}{\includegraphics[scale=0.27]{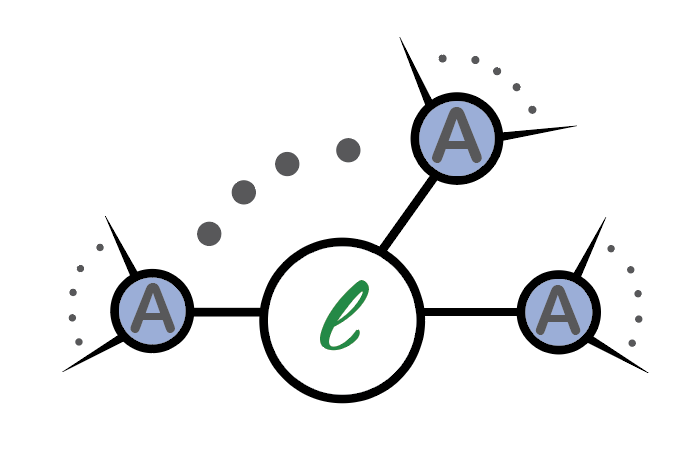}}\xrightarrow[\text{residue}]{\ell^2\rightarrow0 }\sum\parbox{3.cm}{\includegraphics[scale=0.27]{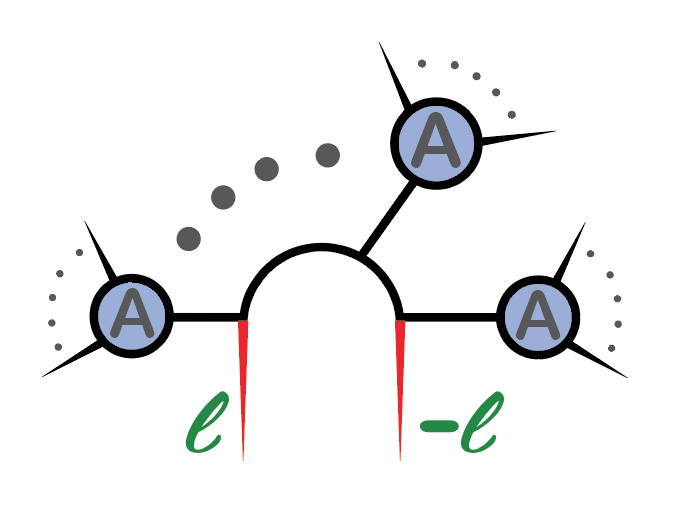}}=A_{\text{tree}}(\pvec_1,...,\lvec,...,-\lvec,...,\pvec_n)\,.
\end{align*}
\noindent\textbf{Dual momenta interlude.} In order to make this happen it is convenient to introduce dual momenta, see \cite{Thorn:2004ie,Chakrabarti:2005ny,Chakrabarti:2006mb}. Any planar diagram gives a number of finite regions that bound loops and a number of infinite region that are bound by external lines extended all the way to infinity. Every region has a dual momentum associated to it. We denote the dual loop momentum $\qvec$ and the dual regional momenta as $\kvec_i$, e.g.  
\begin{align*}
    \includegraphics[trim={0cm 1.0cm 0.0cm 1.0cm},clip,scale=0.3]{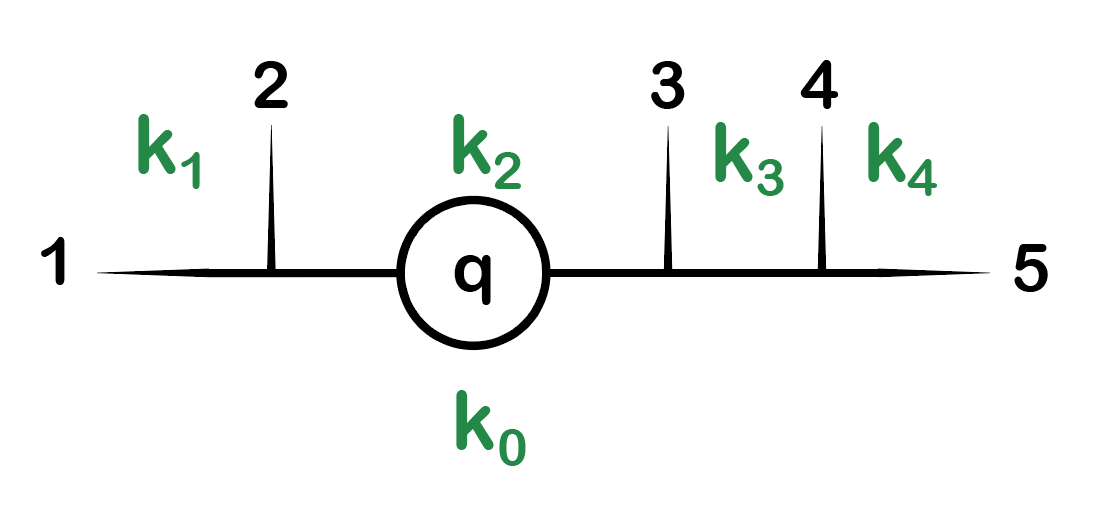}
\end{align*}
Here the relation between the original momenta $\lvec,\pvec_i$ and the dual momenta $\qvec,\kvec_i$ reads
\begin{align}
    \lvec=\qvec-\kvec_0,\qquad \pvec_i=\kvec_i-\kvec_{i-1},\quad \kvec_n\equiv\kvec_0\,.
\end{align}
Each $\kvec_i$ is sandwiched by two external momenta $\pvec_{i+1}$ and $\pvec_i$ while $\qvec$ is shielded by the bubble. Note that for an $n$-point amplitude there are $n$ independent $\kvec_i$ instead of $n-1$ independent $\pvec_i$ (due to momentum conservation). Therefore, there should be a translation symmetry in the dual space to compensate for this redundancy in $\kvec_i$. The physical amplitude must be translation invariant in $\kvec_i$. If this is so, then it is possible to solve for all $\kvec_i$ in terms of external momenta $\pvec_i$. At this point we move to the dual space. Each term in $F$ has a loop and now each segment of the loop has $\qvec-\kvec_i$ flowing through it for a certain $i$. The dual space automatically leads to the correct routing of the momenta. Now, we consider $F$ to be a function of $\qvec$, $\kvec_i$ and are interested in the poles with respect to $\qvec^-$. The residue at each pole gives the sum over all tree level diagrams with the same momenta on the external lines. The latter is crucial for getting the complete tree-level amplitude as the residue (rather than just a random sum of tree-level diagrams with different momenta on some of the external lines). 

\noindent\textbf{Back to the proof.} It turns out that the interactions fine-tuned by the higher spin symmetry make all tree-level amplitudes vanish \cite{Skvortsov:2018jea,Skvortsov:2020wtf}. Therefore, we have a meromorphic function $F$, whose residues vanish. Therefore, $F\equiv 0$. Note that $F$ is just the total one-loop integrand. However, we do not need all terms of $F$ to get the $S$-matrix element. The self-energy corrections should be excluded since they are anomalous (see the discussion after \eqref{eq:2pt}). Also, the tadpoles vanish by themselves. To this end, we represent $F$ as follows
\begin{align}
    F&=F^{1-\text{loop}}_{S} +F^{1-\text{loop}}_{\text{bubbles}}+F^{1-\text{loop}}_{\text{tadpoles}}=0\,,
\end{align}
where $F^{1-\text{loop}}_{S}$ is the complete integrand for the one-loop $S$-matrix element and $F^{1-\text{loop}}_{\text{bubbles}}$, $F^{1-\text{loop}}_{\text{tadpoles}}$ are self-evident. The tadpoles and the cuts of tadpoles vanish by themselves. Indeed, the tadpole has $V(\boldsymbol{0},\mu;\lvec,\lambda;-\lvec,-\lambda)\equiv0$ as a vertex. It is important that the cubic self-interaction of the scalar field is absent, i.e. $V(\pvec_1,0;\pvec_2,0;-\pvec_1-\pvec_2,0)\equiv0$.

There is a nontrivial, but finite, contribution from the self-energy insertions into various external and internal lines, see below. As a result we have
\begin{align}
    F^{1-\text{loop}}_{S} +F^{1-\text{loop}}_{\text{bubbles}}&=0\,, && F^{1-\text{loop}}_{\text{tadpoles}}=0\,.
\end{align}
Therefore, in order to get the full one-loop $S$-matrix element we need to sum over all bubble's insertions. The summation will be done with the help of the tree-level amplitudes that are available \cite{Skvortsov:2018jea,Skvortsov:2020wtf} and we briefly summarize the results. 

\noindent\textbf{Tree-level interlude.} Despite the abundance of interactions among higher spin fields, all tree-level amplitudes can easily be computed \cite{Ponomarev:2016lrm,Skvortsov:2018jea,Skvortsov:2020wtf}. In order to avoid a tedious summation over Feynman diagrams, a variation of the Berends-Giele method can be applied to get a recursion relation. The four-point color-ordered amplitude is given by two diagrams 
\begin{align*}
   \parbox{2.2cm}{\includegraphics[scale=0.3]{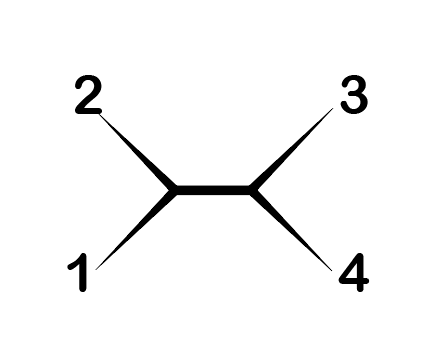}}+\parbox{2.2cm}{\includegraphics[scale=0.3]{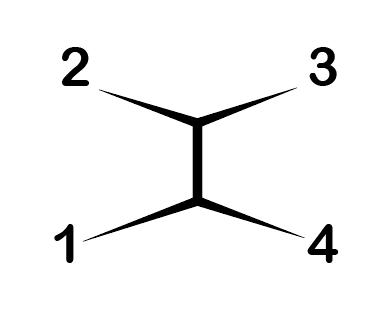}}
\end{align*}
which leads to \cite{Ponomarev:2016lrm,Skvortsov:2018jea,Skvortsov:2020wtf}, 
\begin{equation}
    A_4(1234)=\frac{\delta(\sum_i\pvec_i)}{\Gamma(\Lambda_4-1)\prod_{i=1}^4\beta_i^{\lambda_i}}\Big[\frac{\PPb_{12}\PPb_{34}(\PPb_{12}+\PPb_{34})^{\Lambda_4-2}}{(\pvec_1+\pvec_2)^2}+\frac{\PPb_{23}\PPb_{41}(\PPb_{23}+\PPb_{41})^{\Lambda_4-2}}{(\pvec_2+\pvec_3)^2}\Big]\,,
\end{equation}
where $\Lambda_4=\lambda_1+...+\lambda_4$ and we drop the overall $\delta$-function in what follows. Setting all legs save for the first one on-shell, $\pvec_1^2\neq0$, with a help of simple kinematical identities, it can be simplified to
\begin{equation}
    A_4(1234)=\frac{\alpha_4^{\Lambda_4-2}}{\Gamma(\Lambda_4-1)\prod_{i=1}^4\beta_i^{\lambda_i-1}}\frac{\beta_3\, \pvec_1^2}{4\beta_1\PP_{23}\PP_{34}}\,,
\end{equation}
where $\alpha_4=\PPb_{12}+\PPb_{34}=\PPb_{23}+\PPb_{41}$ is cyclic invariant. The recursion results in the following $n$-point amplitude 
\begin{equation}\label{eq:npointrecursive}
    A_n(1...n)=\frac{(-)^n\,\alpha_n^{\Lambda_n-(n-2)}\beta_3...\beta_{n-1}\,\pvec_1^2}{2^{n-2}\Gamma(\Lambda_n-(n-3))\prod_{i=1}^{n}\beta_i^{\lambda_i-1}\beta_1\PP_{23}...\PP_{n-1,n}}\,, 
    \end{equation}
    \begin{equation} \label{alpha-n}
    \alpha_n=\sum_{i<j}^{n-2}\PPb_{ij}+\PPb_{n-1,n}\,,
\end{equation}
where $\Lambda_n=\lambda_1+...+\lambda_n$. As is claimed, it vanishes on-shell, which is manifested by the overall $\pvec_1^2$ factor. Consistently with the Weinberg and Coleman-Mandula theorems the tree-level amplitudes have to vanish on-shell in any higher spin gravity in flat space (they do not vanish off-shell). This raises an interesting question of what are the possible observables in a higher spin gravity that are Lorentz, hence higher spin, invariant and do not vanish on-shell, if any. We note, that $S=1$ in flat space has an AdS/CFT counterpart \cite{Klebanov:2002ja,Sezgin:2002rt,Leigh:2003gk,Maldacena:2011jn,Boulanger:2013zza} where the higher spin invariant holographic $S$-matrix has to be given by a free CFT's correlation functions which are the simplest invariants of higher spin symmetry \cite{Colombo:2012jx,Didenko:2013bj,Didenko:2012tv,Bonezzi:2017vha}.  

\noindent\textbf{Self-energy interlude.} One more ingredient is the self-energy correction, which is a subtle diagram to compute even for QCD in the light-cone gauge \cite{Chakrabarti:2005ny,Chakrabarti:2006mb}. We stress that we will use dual momenta. The diagram is regularized with the help of a Gaussian cut-off $\exp [-\xi q_\perp^2]$ in the transverse part of $\qvec$. The diagram turns out to be UV-convergent, but one has to analyze carefully the $\xi  \rightarrow 0$ limit,  \cite{Chakrabarti:2005ny,Skvortsov:2020wtf}. The result for the  self-energy diagram in the planar limit of the $U(N)$-gauged Chiral Theory or for the $N=1$ theory is \cite{Skvortsov:2018jea,Skvortsov:2020wtf}
\begin{equation}\label{eq:2pt}
\begin{split}
    \parbox{3.8cm}{\includegraphics[trim={0cm 0.3cm 0.0cm 0.6cm},clip,scale=0.38]{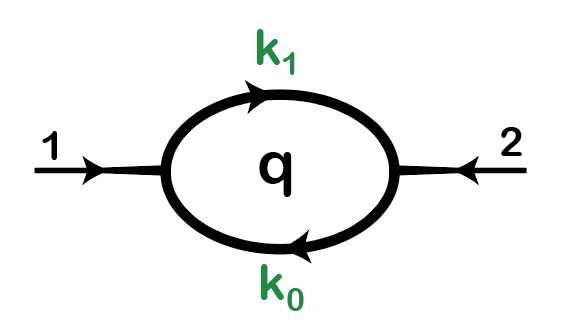}}&=\nu_0\,N\frac{(l_p)^{\Lambda_2-2}}{\beta_1^{\lambda_1}\beta_2^{\lambda_2}\Gamma[\Lambda_2-1]}\int \frac{d^4q}{(2\pi)^4} \frac{\PPb_{q-k_0,p_1}^2\delta_{\Lambda_2,2}}{(\qvec-\kvec_0)^2(\qvec-\kvec_1)^2}\\
    &=\nu_0 N(\kb_0^2+\kb_0\kb_1+\kb_1^2)\frac{\delta_{\Lambda_2,2}(l_p)^{\Lambda_2-2}}{96\pi^2\beta_1^{\lambda_1-1}\beta_2^{\lambda_2-1}\Gamma[\Lambda_2-1]}\,,
    \end{split}
\end{equation}
where $\nu_0=\sum_{\lambda} 1$. It is important to note that the result is non-vanishing only when $\Lambda_2=\lambda_1+\lambda_2=2$. Below, we set Planck's length $l_p=1$ for simplicity. There is an overall numerical factor $\nu_0$. It results from the summation over the helicities running in the loop. There are two such sums, since there are two segments of the loop, and the second one factors out after the first one is evaluated. Clearly, $\nu_0$ counts the number of degrees of freedom in the theory and has nothing to do with the UV-convergence. We will discuss $\nu_0$ at the end. Note that the amplitude is not translation invariant in the dual space, i.e. it is anomalous. Therefore, it has to removed by a counterterm, which will be important later.

\noindent\textbf{Inserting bubbles into tree-level diagrams.} As for the tree-level amplitudes, the direct summation over all tree-level diagrams with the bubble inserted is hardly feasible. Instead, in order to compute $F^{1-\text{loop}}_{\text{bubbles}}$ we apply another recursive relation, which can be depicted as 
\begin{align}\label{eq:recursivegraph}
   \sum_{i=1}^{[n/2]} \parbox{1.8cm}{\includegraphics[scale=0.2]{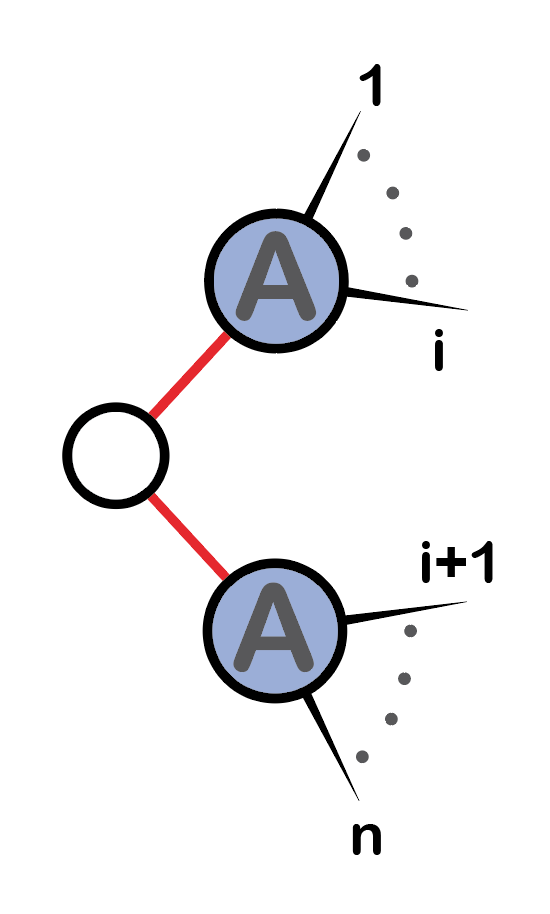}}=\Bigg[\parbox{2.82cm}{\includegraphics[scale=0.27]{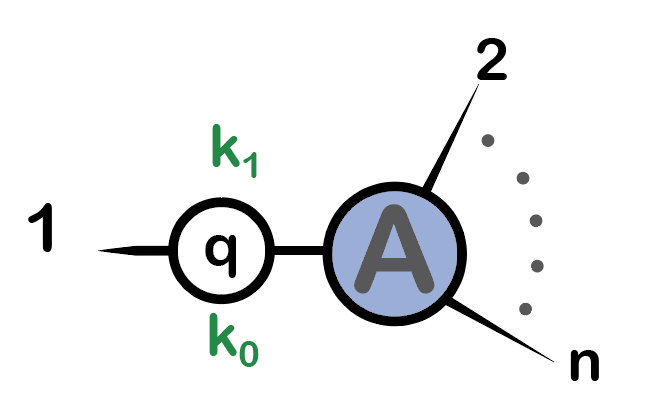}}+\parbox{2.88cm}{\includegraphics[scale=0.27]{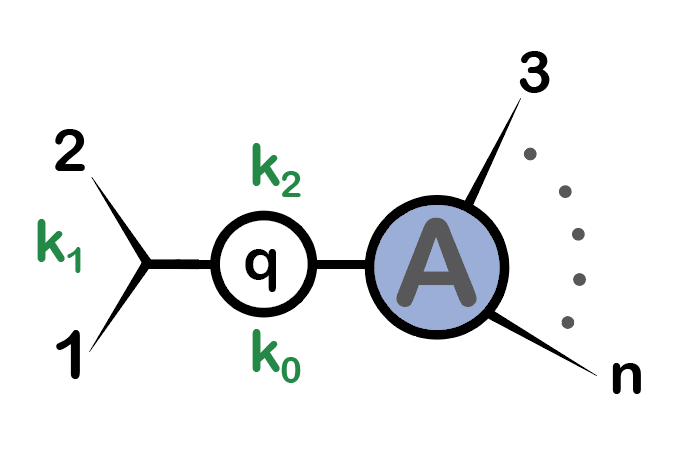}}+\parbox{3.6cm}{\includegraphics[scale=0.265]{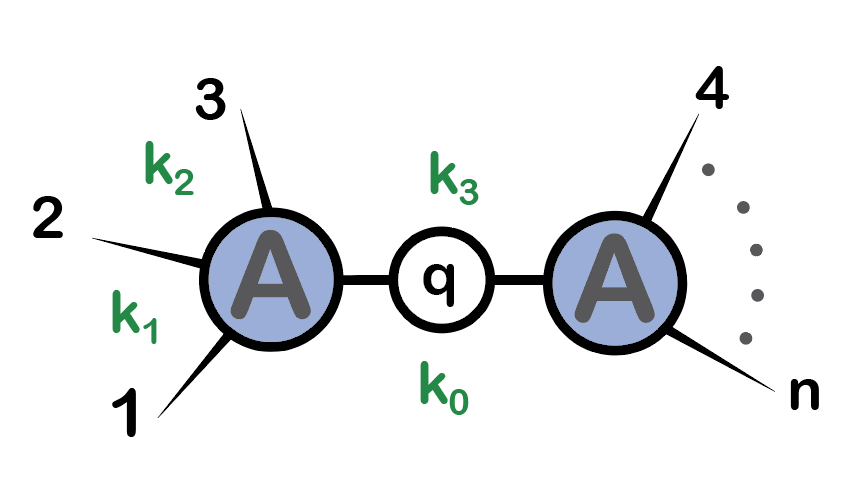}}+...\Bigg]\,.
\end{align}
Here, the blue blobs are the tree-level sub-amplitudes that are being glued to the bubble (the white blob). First, the white blob sits on the leftmost external line. In the second term it is one vertex away from the external lines on the left. In the third term it has passed three external lines on the left and so on. The final $\ldots$ also implies the sum over the cyclic permutations. Inserting the self-energy integral \eqref{eq:2pt} will give a contribution of
\begin{align}\label{eq:qintegral}
    &(k_j^+-k_i^+)^2(\kb_i^2+\kb_i\kb_j+\kb_j^2)\,,&& k_j^+-k_i^+=\sum_{m=i+1}^{j}\beta_m\,,
\end{align}
where $\kvec_{i,j}$ are the regional dual momenta that are adjacent to the inserted bubble. Note that once we insert the bubble into an internal line, the two propagators get cancelled against the $\pvec^2$-factors of the two tree-level diagrams \eqref{eq:npointrecursive} being glued. We also note that the bubble is slightly off-diagonal in the helicity space since it has $\delta_{\lambda_1+\lambda_2,2}$ instead of $\delta_{\lambda_1+\lambda_2,0}$ for the propagators. 

\noindent\textbf{One-loop amplitude.} What remains is to massage the sum over the bubble's insertions and to put the minus sign in front. Let us write \eqref{eq:recursivegraph} in terms of $\PP,\kb$ and $\beta$ components by using \eqref{eq:npointrecursive} and \eqref{eq:qintegral}. The diagrams in \eqref{eq:recursivegraph} correspond to gluing the bubble to the two sub-amplitudes with the total number of external legs equal $n$ and then taking the cyclic permutations. We arrive at
\small
\begin{equation}\label{eq:recursiveexplicit}
\begin{split}
    (2.10)&=\mathcal{N}_n\Bigg[\sum_{i=1}^{[n/2]}\frac{(\kb_0^2+\kb_0\kb_i+\kb_i^2)(\sum_{k=1}^i\beta_k)^2}{\beta_1\beta_i\beta_{i+1}\beta_n}\PP_{i,i+1}\PP_{n1}+\text{cyclic permutations}\Bigg]
    \end{split}
\end{equation}
\normalsize
where 
\begin{align}
    \mathcal{N}_n=\nu_0\,\frac{(-1)^{n}\alpha_n^{\Lambda_n-n}}{2^{n+3}3\pi^2\Gamma[\Lambda_n-(n-1)]\prod_{i=1}^n\beta_i^{\lambda_i-2}\PP_{12}\PP_{23}...\PP_{n1}}\,.
\end{align}
As we have already stressed, all physical quantities must be translation invariant in the dual space. Therefore, \eqref{eq:recursiveexplicit} should not change if we replace $\kvec_i$ by $\kvec_i+\boldsymbol{a}$ for any $\boldsymbol{a}$. One way to see it is to solve for all $\kvec_i$ except for $\kvec_0$ via $\kvec_i=\kvec_0+\sum_{j=1}^i\pvec_j$. In order to see that the resulting expression $f(\kb_0)$ does not depend on $\kb_0$ we can take its derivative $f'(\kb_0)$ to get
\begin{align}
    \mathcal{N}_n\kb_0&\Bigg[\sum_{i=1}^{[n/2]}\frac{(\sum_{k=1}^i\beta_k)^2}{\beta_1\beta_i\beta_{i+1}\beta_n}\PP_{i,i+1}\PP_{n1}+\text{cyclic permutations}\Bigg]\,.
\end{align}
This is nothing but \eqref{eq:recursiveexplicit} with all $(\kb_i^2+\kb_i\kb_j+\kb_j^2)$ factors erased, times $\kb_0$. It is easy to show that this expression is indeed zero with the help of the momentum conservation, see various identities in \cite{Skvortsov:2020wtf}. Once \eqref{eq:recursiveexplicit} is shown to be translation invariant, it can be expressed in terms of external momenta $\pvec_i$ only. This is quite remarkable since the self-energy diagram itself, \eqref{eq:2pt}, is not translation invariant, it is anomalous. 

Due to many kinematical identities involving $\beta_i$ and $\PPb_{ij}$, there is no unique way to write the final result, but the following form is very suggestive
\begin{align}\label{eq:result}
    A_{\text{1-loop}}^{\text{HSG}}=\Big[\sum_{1\leq i_1<i_2<i_3<i_4\leq n}\frac{ \check{\beta}^{n-4}_{i_1,i_2,i_3,i_4}\PP_{i_1i_2}\PPb_{i_2i_3}\PP_{i_3i_4}\PPb_{i_4i_1}}{2^{\tfrac{n}2-2}\PP_{12}\PP_{23}...\PP_{n1}}\Big]\times D^{\text{HSG}} \times\nu_0\,,
\end{align}
where 
\begin{align}
    \check{\beta}^{n-4}_{i_1,i_2,i_3,i_4}&=\frac{\prod_{j=1}^{n}\beta_j}{\beta_{i_1}\beta_{i_2}\beta_{i_3}\beta_{i_4}}\,, & D^{\text{HSG}}&=\frac{(-)^{n}\alpha_n^{\Lambda_n-n}}{2^{\frac{n}{2}+5}3\pi^2\Gamma[\Lambda_n-(n-1)]\prod_{i=1}^n\beta_i^{\lambda_i-1}}\,.
\end{align}
Clearly, the one-loop amplitude in Chiral Higher Spin Gravity consists of (i) a factor that has a lower spin origin as it does not have enough $\PP$ to account for $\lambda_i$; (ii) kinematical higher spin dressing factor $D^{\text{HSG}}$ that accounts for helicities $\lambda_i$ on the external lines, which the first factor cannot accomplish; (iii) the total number of physical degrees of freedom $\nu_0$. 

The first factor is telling. Applying the light-cone vs. spinor-helicity dictionary \eqref{9a1}, we discover the all-plus helicity one-loop amplitude in QCD or in self-dual Yang-Mills \cite{Bern:1993sx,Bern:1993qk,Mahlon:1993fe}:
\begin{align}
    A_{\text{SDYM, 1-loop}}=A_{\text{QCD,1-loop}}^{++...+}=\sum_{1\leq i_1<i_2<i_3<i_4\leq n}\frac{\langle i_1i_2\rangle[i_2i_3]\langle i_3i_4\rangle[i_4i_1]}{\langle 12\rangle \langle 23 \rangle ... \langle n1\rangle}\,.
\end{align}
In other words, the one-loop amplitude in Chiral Higher Spin Gravity is found to be
\begin{align}\label{finalres}
    A_{\text{1-loop}}^{\text{HSG}} =A_{\text{QCD,1-loop}}^{++...+}\times D^{\text{HSG}} \times \nu_0\,.
\end{align}
Therefore, we get precisely the structure \eqref{result} that is sketched in the introduction. Moreover, when we set $\lambda_i=1$, \eqref{finalres} reduces to just the SDYM/QCD amplitude times an overall numerical factor, i.e. the higher spin dressing disappears. 

\noindent\textbf{Vacuum one-loop bubbles postlude.} The last step is to explain what to do with the overall numerical factor $\nu_0$. Any theory with infinitely many fields is not quite a field theory in at least one respect: a prescription must be given on how to sum over the spectrum. This issue is clearly visible in Kaluza-Klein reduction, see e.g. \cite{Fradkin:1982kf}, and Exceptional Field Theories \cite{Bossard:2015foa}. Due to the specific structure of interactions the same issue is not faced in Chiral Theory at the tree level. Also, at the loop level, the only problem is to assign some value to just one numerical sum. The situation is more complicated in conformal HSG \cite{Joung:2015eny} and in a more general $AdS_4$-HSG \cite{Bekaert:2015tva} (if the latter exists as a field theory, then Chiral HSG is a subsector therein), where infinite sums occur already at tree-level.

We would like to review that the infinite sum $\nu_0$ is just the simplest example out of a large web of results on one-loop determinants in higher spin theories \cite{Gopakumar:2011qs,Tseytlin:2013jya,Giombi:2013fka,Giombi:2014yra,Beccaria:2014jxa,Beccaria:2014xda,Beccaria:2015vaa,Gunaydin:2016amv,Bae:2016rgm,Skvortsov:2017ldz}. In the absence of a theory, but having some prediction for its spectrum, one can compute a vacuum one-loop partition function. The action of a free higher spin gravity reads
\begin{align*}
    S&= \sum_s \int \Phi_{s} \mathrm{K}_s \Phi_{s}\,, && \mathrm{K}_s=-\nabla^2+M_s^2\,, && \delta\Phi_{a_1...a_s}=\nabla_{a_1} \xi_{a_2...a_s}+\text{perm}\,,
\end{align*}
where we partially fixed the gauge symmetry, so that the kinetic operator is just the Laplace operator on Minkowski or anti-de Sitter space. We appeal to the covariant formulation for a moment, where a spin-$s$ field is described by a symmetric rank-$s$ tensor. The gauge transformations are also displayed. The one-loop partition function is
\begin{align}\label{oneloopdet}
    e^{-2F_{}}=Z^2&=\frac{1}{\det \mathrm{K}_0}\frac{ \det \tilde{\mathrm{K}}_0 \det\tilde{\mathrm{K}}_1...} { \det\mathrm{K}_1 \det\mathrm{K}_2...}
\end{align}
where $\tilde{\mathrm{K}}_s$ correspond to the kinetic terms of ghosts. To compute the free energy $F$ in anti-de Sitter space is technically challenging for two reasons: the spectral zeta function of the Laplace operator on transverse traceless rank-$s$ tensor is quite complicated. In addition the infinite sum/product over all species needs to be taken, which requires a clever regularization. We summarize the results for the free energy $F$ in four dimensions \cite{Giombi:2013fka,Beccaria:2015vaa} below
\begin{center}
\begin{tabular}{c|c|c}
            & all integer spins & all even spins\\ \hline
flat space  &  $\sum_\lambda 1=0$ & $\sum_{2\lambda} 1=0$\\
$AdS_4\,, \Delta=1$     &  0 &   $\frac{1}{16} \left(\log (4)-\frac{3 \zeta (3)}{\pi ^2}\right)$ \\
$AdS_4\,, \Delta=2$     & $-\displaystyle\frac{\zeta (3)}{8 \pi ^2}$ &  $\frac{1}{16} \left(\log (4)-\frac{5 \zeta (3)}{\pi ^2}\right)$
\end{tabular}
\end{center}
where $\Delta=1,2$ corresponds to the boundary conditions on the scalar field $\Phi_0$. The $AdS_4$ results are highly nontrivial and the numbers fit quite well the conjectured $AdS_4/CFT_3$ dualities \cite{Giombi:2013fka}. In the Minkowski case \cite{Beccaria:2015vaa} there are no mass-like terms in $\mathrm{K}_s$, $\tilde{\mathrm{K}}_s$ and, as a result $\mathrm{K}_s=\tilde{\mathrm{K}}_s$. Therefore, in \eqref{oneloopdet} it is tempting to cancel the denominators against the corresponding terms in the numerator at least when the theory has all integer spins. This gives $Z=1$ and $F=0$. Note that the $AdS_4$ result for $\Delta=1$, i.e. for the free vector model on the boundary, also gives $F=0$, but this occurs as a result of a nontrivial cancellation over all spins. Since $F$ counts the number of effective degrees of freedom, another way to ensure $F=0$ is to properly count the states:
\begin{align}\label{pdof}
    \nu_0=\sum_{\lambda} 1= 1+2 \sum_{\lambda>0}\lambda=1+2\zeta(0)=0\,,
\end{align}
i.e. each massless field contributes two degrees of freedom and the scalar contributes one degree of freedom. To conclude, both the Weinberg, Coleman-Mandula theorems and the one-loop determinants instruct us to set $\nu_0=0$ and get $S=1$. This can safely be done since the one-loop amplitude is shown to be UV-finite.  

\section{Conclusions and Discussion}
\label{sec:conclusions}
We have shown that Chiral Higher Spin Gravity (HSG) is one-loop finite in flat space. This is encouraging and indicates that the higher spin programme has good chances to fulfill its goal to construct viable models of quantum gravity. Our result guarantees that the leading divergences go away in Chiral Theory on $AdS_4$ as well. There are, however, certain lower derivative tails present in the $AdS$-uplift of the light-cone vertices \cite{Metsaev:2018xip} and it has to be checked whether these additional terms can generate any UV-divergences.

$4d$ Gravity is known to be one-loop finite: thanks to the Gauss-Bonnet invariant the only seemingly non-vanishing counterterm $C_{\mu\nu,\lambda\rho}^2$ can be reduced to $R^2$ and $R_{\mu\nu}^2$ that vanish on-shell. As a result, there is no need to check one-loop diagrams with arbitrary number of external legs. There is no similar argument for Chiral Theory or any other HSG. The number of derivatives grows with spin. Therefore, one-loop diagrams with any number of legs are naively UV-divergent and need to be examined, which is what we did in the paper.

Chiral Theory is a subsector of any other HSG in $AdS_4$ that has the same spectrum. Therefore, our results ensure cancellation of UV-divergences in a large class of diagrams in these more general theories. However, there are certain new quantum corrections that correspond to the chiral and anti-chiral interactions appearing in the same diagram or coming from the other vertices that are neither chiral nor anti-chiral. The latter vertices are known to be non-local \cite{Bekaert:2015tva,Maldacena:2015iua,Sleight:2017pcz,Ponomarev:2017qab}, with the non-locality going beyond the one allowed by the field theory approach. It remains to be seen if the non-localities can be tamed first at the classical and then at the quantum levels. Note that in Chiral Theory the infinite sum over the states shows up at the loop level only and has a simple form of a numerical factor, but in more general HSGs, which have both chiral and anti-chiral interactions, the problem appears already at the tree-level and it is not about an overall factor. The collision of the chiral and anti-chiral sectors is at the core of the non-locality problem.  

There are two more encouraging quantum checks of HSG that have been done so far: (i) one-loop determinants that we already mentioned \cite{Gopakumar:2011qs,Tseytlin:2013jya,Giombi:2013fka,Giombi:2014yra,Beccaria:2014jxa,Beccaria:2014xda,Beccaria:2015vaa,Gunaydin:2016amv,Bae:2016rgm,Skvortsov:2017ldz}; (ii) one-loop corrections to the four-point function via AdS unitarity cuts \cite{Ponomarev:2019ltz,Ponomarev:2019ofr}. The first one probes the free spectrum of states. The second one reduces the problem of one-loop corrections to a certain manipulation with the CFT data of the dual CFT, while a direct check would be beneficial. 

An interesting feature of the obtained results is a relation to the self-dual subsector of QCD, which is consistent with \cite{Ponomarev:2017nrr}. It is quite unexpected that a higher spin gravity theory has any simple relation to the real world non-gravitational physics. At the same time the holographic $S$-matrix of Chiral Theory on $AdS_4$ is directly related to Chern-Simons Matter theories. At the level of three-point functions it allows to determine correlation functions of higher spin currents in the large-$N$ Chern-Simons Matter theories for any value of the 't Hooft coupling \cite{Skvortsov:2018uru}. More generally, Chiral HSG should be dual to a certain subsector of Chern-Simons Matter theories. It would be interesting to see how far the relation to the QCD amplitudes extends into the HSG world and into Chern-Simons Matter theories.

As our proof indicates, the cancellation of UV-divergences in HSGs is a very subtle effect. Indeed, (at least some of) the one-loop amplitudes are related to the all-plus helicity amplitudes in QCD and SDYM. These amplitudes are quite sophisticated: the naive integrand vanishes in $d=4$ and in dimensional regularization it can be seen to vanish as $(d-4)$, while the loop integral gives a simple pole $(d-4)^{-1}$, the final result being finite. In the light-cone gauge the appearance is different: the entire one-loop amplitude comes from the insertions of the self-energy correction into tree-level diagrams. The self-energy correction turns out to be finite, but it violates the translation symmetry in the dual momentum space. Therefore, it is anomalous. Nevertheless, the sum over all self-energy insertions turns out to be translation invariant and, hence, well-defined. There is a lot of similarities between HSGs and topological theories (e.g. the effective number of degrees of freedom vanishes in the topologically trivial setup), like Chern-Simons theory, where nontrivial effects arise due to nontrivial topology and/or require a subtle regularization.

It would very interesting to explore further the relation between Chiral HSG and self-dual Yang-Mills theory \cite{Ponomarev:2017nrr}.  For example, the two- and higher-loop amplitudes should vanish identically. Therefore, Chiral Higher Spin Theory should be one-loop exact.

\section*{Acknowledgments}
\label{sec:Aknowledgements}
We would like to thank  Gregory Korchemsky, Ruslan Metsaev, Jan Plefka, Dmitry Ponomarev, Charles Thorn, Arkady Tseytlin and Mirian Tsulaia for useful discussions and comments.  The work of E.S. was supported by the Russian Science Foundation grant 18-72-10123 in association with the Lebedev Physical Institute. The work of T.T. is supported by the International Max Planck Research School for Mathematical and Physical Aspects of Gravitation, Cosmology and Quantum Field Theory.

\setstretch{1.0}
\footnotesize
\providecommand{\href}[2]{#2}\begingroup\raggedright\endgroup


\begin{thebibliography}{10}

\bibitem{Metsaev:1991mt}
R.~R. Metsaev, {\it {Poincare invariant dynamics of massless higher spins:
  Fourth order analysis on mass shell}},  {\em Mod. Phys. Lett.} {\bf A6}
  (1991) 359--367.

\bibitem{Metsaev:1991nb}
R.~R. Metsaev, {\it {S matrix approach to massless higher spins theory. 2: The
  Case of internal symmetry}},  {\em Mod. Phys. Lett.} {\bf A6} (1991)
  2411--2421.

\bibitem{Ponomarev:2016lrm}
D.~Ponomarev and E.~D. Skvortsov, {\it {Light-Front Higher-Spin Theories in
  Flat Space}},  {\em J. Phys.} {\bf A50} (2017), no.~9 095401
  [\href{http://arXiv.org/abs/1609.04655}{{\tt 1609.04655}}].

\bibitem{Skvortsov:2018jea}
E.~D. Skvortsov, T.~Tran and M.~Tsulaia, {\it {Quantum Chiral Higher Spin
  Gravity}},  {\em Phys. Rev. Lett.} {\bf 121} (2018), no.~3 031601
  [\href{http://arXiv.org/abs/1805.00048}{{\tt 1805.00048}}].

\bibitem{Skvortsov:2020wtf}
E.~Skvortsov, T.~Tran and M.~Tsulaia, {\it {More on Quantum Chiral Higher Spin
  Gravity}},  \href{http://arXiv.org/abs/2002.08487}{{\tt 2002.08487}}.

\bibitem{Fronsdal:1978rb}
C.~Fronsdal, {\it Massless fields with integer spin},  {\em Phys. Rev.} {\bf
  D18} (1978) 3624.

\bibitem{Fradkin:1986ka}
E.~S. Fradkin and M.~A. Vasiliev, {\it Candidate to the role of higher spin
  symmetry},  {\em Ann. Phys.} {\bf 177} (1987) 63.

\bibitem{Flato:1978qz}
M.~Flato and C.~Fronsdal, {\it {One Massless Particle Equals Two Dirac
  Singletons: Elementary Particles in a Curved Space. 6.}},  {\em
  Lett.Math.Phys.} {\bf 2} (1978) 421--426.

\bibitem{Berends:1984wp}
F.~A. Berends, G.~J.~H. Burgers and H.~Van~Dam, {\it {On spin three
  selfinteractions}},  {\em Z. Phys.} {\bf C24} (1984) 247--254.

\bibitem{Maldacena:2011jn}
J.~Maldacena and A.~Zhiboedov, {\it {Constraining Conformal Field Theories with
  A Higher Spin Symmetry}},  {\em J. Phys.} {\bf A46} (2013) 214011
  [\href{http://arXiv.org/abs/1112.1016}{{\tt 1112.1016}}].

\bibitem{Boulanger:2013zza}
N.~Boulanger, D.~Ponomarev, E.~D. Skvortsov and M.~Taronna, {\it {On the
  uniqueness of higher-spin symmetries in AdS and CFT}},  {\em Int. J. Mod.
  Phys.} {\bf A28} (2013) 1350162 [\href{http://arXiv.org/abs/1305.5180}{{\tt
  1305.5180}}].

\bibitem{Weinberg:1964ew}
S.~Weinberg, {\it {Photons and Gravitons in s Matrix Theory: Derivation of
  Charge Conservation and Equality of Gravitational and Inertial Mass}},  {\em
  Phys. Rev.} {\bf 135} (1964) B1049--B1056.

\bibitem{Coleman:1967ad}
S.~R. Coleman and J.~Mandula, {\it {All Possible Symmetries of the S Matrix}},
  {\em Phys. Rev.} {\bf 159} (1967) 1251--1256.

\bibitem{Fotopoulos:2010ay}
A.~Fotopoulos and M.~Tsulaia, {\it {On the Tensionless Limit of String theory,
  Off - Shell Higher Spin Interaction Vertices and BCFW Recursion Relations}},
  {\em JHEP} {\bf 11} (2010) 086 [\href{http://arXiv.org/abs/1009.0727}{{\tt
  1009.0727}}].

\bibitem{Bekaert:2010hp}
X.~Bekaert, N.~Boulanger and S.~Leclercq, {\it {Strong obstruction of the
  Berends-Burgers-van Dam spin-3 vertex}},  {\em J. Phys.} {\bf A43} (2010)
  185401 [\href{http://arXiv.org/abs/1002.0289}{{\tt 1002.0289}}].

\bibitem{Roiban:2017iqg}
R.~Roiban and A.~A. Tseytlin, {\it {On four-point interactions in massless
  higher spin theory in flat space}},  {\em JHEP} {\bf 04} (2017) 139
  [\href{http://arXiv.org/abs/1701.05773}{{\tt 1701.05773}}].

\bibitem{Dempster:2012vw}
P.~Dempster and M.~Tsulaia, {\it {On the Structure of Quartic Vertices for
  Massless Higher Spin Fields on Minkowski Background}},  {\em Nucl. Phys.}
  {\bf B865} (2012) 353--375 [\href{http://arXiv.org/abs/1203.5597}{{\tt
  1203.5597}}].

\bibitem{Bekaert:2015tva}
X.~Bekaert, J.~Erdmenger, D.~Ponomarev and C.~Sleight, {\it {Quartic AdS
  Interactions in Higher-Spin Gravity from Conformal Field Theory}},  {\em
  JHEP} {\bf 11} (2015) 149 [\href{http://arXiv.org/abs/1508.04292}{{\tt
  1508.04292}}].

\bibitem{Maldacena:2015iua}
J.~Maldacena, D.~Simmons-Duffin and A.~Zhiboedov, {\it {Looking for a bulk
  point}},  {\em JHEP} {\bf 01} (2017) 013
  [\href{http://arXiv.org/abs/1509.03612}{{\tt 1509.03612}}].

\bibitem{Sleight:2017pcz}
C.~Sleight and M.~Taronna, {\it {Higher spin gauge theories and bulk locality:
  a no-go result}},  \href{http://arXiv.org/abs/1704.07859}{{\tt 1704.07859}}.

\bibitem{Ponomarev:2017qab}
D.~Ponomarev, {\it {A Note on (Non)-Locality in Holographic Higher Spin
  Theories}},  {\em Universe} {\bf 4} (2018), no.~1 2
  [\href{http://arXiv.org/abs/1710.00403}{{\tt 1710.00403}}].

\bibitem{Fradkin:1986qy}
E.~S. Fradkin and M.~A. Vasiliev, {\it {Cubic Interaction in Extended Theories
  of Massless Higher Spin Fields}},  {\em Nucl. Phys.} {\bf B291} (1987) 141.

\bibitem{Conde:2016izb}
E.~Conde, E.~Joung and K.~Mkrtchyan, {\it {Spinor-Helicity Three-Point
  Amplitudes from Local Cubic Interactions}},  {\em JHEP} {\bf 08} (2016) 040
  [\href{http://arXiv.org/abs/1605.07402}{{\tt 1605.07402}}].

\bibitem{Bekaert:2010hw}
X.~Bekaert, N.~Boulanger and P.~Sundell, {\it {How higher-spin gravity
  surpasses the spin two barrier: no-go theorems versus yes-go examples}},
  {\em Rev.Mod.Phys.} {\bf 84} (2012) 987--1009
  [\href{http://arXiv.org/abs/1007.0435}{{\tt 1007.0435}}].

\bibitem{Bengtsson:1983pd}
A.~K.~H. Bengtsson, I.~Bengtsson and L.~Brink, {\it {Cubic Interaction Terms
  for Arbitrary Spin}},  {\em Nucl. Phys.} {\bf B227} (1983) 31--40.

\bibitem{Bengtsson:1983pg}
A.~K.~H. Bengtsson, I.~Bengtsson and L.~Brink, {\it {Cubic Interaction Terms
  for Arbitrarily Extended Supermultiplets}},  {\em Nucl. Phys.} {\bf B227}
  (1983) 41--49.

\bibitem{Blencowe:1988gj}
M.~Blencowe, {\it {A Consistent Interacting Massless Higher Spin Field Theory
  in $D$ = (2+1)}},  {\em Class.Quant.Grav.} {\bf 6} (1989) 443.

\bibitem{Bergshoeff:1989ns}
E.~Bergshoeff, M.~P. Blencowe and K.~S. Stelle, {\it {Area Preserving
  Diffeomorphisms and Higher Spin Algebra}},  {\em Commun. Math. Phys.} {\bf
  128} (1990) 213.

\bibitem{Campoleoni:2010zq}
A.~Campoleoni, S.~Fredenhagen, S.~Pfenninger and S.~Theisen, {\it {Asymptotic
  symmetries of three-dimensional gravity coupled to higher-spin fields}},
  {\em JHEP} {\bf 1011} (2010) 007 [\href{http://arXiv.org/abs/1008.4744}{{\tt
  1008.4744}}].

\bibitem{Henneaux:2010xg}
M.~Henneaux and S.-J. Rey, {\it {Nonlinear $W_{infinity}$ as Asymptotic
  Symmetry of Three-Dimensional Higher Spin Anti-de Sitter Gravity}},  {\em
  JHEP} {\bf 12} (2010) 007 [\href{http://arXiv.org/abs/1008.4579}{{\tt
  1008.4579}}].

\bibitem{Pope:1989vj}
C.~N. Pope and P.~K. Townsend, {\it {Conformal Higher Spin in
  (2+1)-dimensions}},  {\em Phys. Lett.} {\bf B225} (1989) 245--250.

\bibitem{Fradkin:1989xt}
E.~S. Fradkin and V.~{\relax Ya}. Linetsky, {\it {A Superconformal Theory of
  Massless Higher Spin Fields in $D$ = (2+1)}},  {\em Mod. Phys. Lett.} {\bf
  A4} (1989) 731. [Annals Phys.198,293(1990)].

\bibitem{Grigoriev:2019xmp}
M.~Grigoriev, I.~Lovrekovic and E.~Skvortsov, {\it {New Conformal Higher Spin
  Gravities in $3d$}},  {\em JHEP} {\bf 01} (2020) 059
  [\href{http://arXiv.org/abs/1909.13305}{{\tt 1909.13305}}].

\bibitem{Segal:2002gd}
A.~Y. Segal, {\it {Conformal higher spin theory}},  {\em Nucl. Phys.} {\bf
  B664} (2003) 59--130 [\href{http://arXiv.org/abs/hep-th/0207212}{{\tt
  hep-th/0207212}}].

\bibitem{Tseytlin:2002gz}
A.~A. Tseytlin, {\it {On limits of superstring in AdS(5) x S**5}},  {\em Theor.
  Math. Phys.} {\bf 133} (2002) 1376--1389
  [\href{http://arXiv.org/abs/hep-th/0201112}{{\tt hep-th/0201112}}]. [Teor.
  Mat. Fiz.133,69(2002)].

\bibitem{Bekaert:2010ky}
X.~Bekaert, E.~Joung and J.~Mourad, {\it {Effective action in a higher-spin
  background}},  {\em JHEP} {\bf 02} (2011) 048
  [\href{http://arXiv.org/abs/1012.2103}{{\tt 1012.2103}}].

\bibitem{Das:2003vw}
S.~R. Das and A.~Jevicki, {\it {Large N collective fields and holography}},
  {\em Phys. Rev.} {\bf D68} (2003) 044011
  [\href{http://arXiv.org/abs/hep-th/0304093}{{\tt hep-th/0304093}}].

\bibitem{deMelloKoch:2018ivk}
R.~de~Mello~Koch, A.~Jevicki, K.~Suzuki and J.~Yoon, {\it {AdS Maps and
  Diagrams of Bi-local Holography}},  {\em JHEP} {\bf 03} (2019) 133
  [\href{http://arXiv.org/abs/1810.02332}{{\tt 1810.02332}}].

\bibitem{Sperling:2017dts}
M.~Sperling and H.~C. Steinacker, {\it {Covariant 4-dimensional fuzzy spheres,
  matrix models and higher spin}},  {\em J. Phys.} {\bf A50} (2017), no.~37
  375202 [\href{http://arXiv.org/abs/1704.02863}{{\tt 1704.02863}}].

\bibitem{Metsaev:2018xip}
R.~R. Metsaev, {\it {Light-cone gauge cubic interaction vertices for massless
  fields in AdS(4)}},  {\em Nucl. Phys.} {\bf B936} (2018) 320--351
  [\href{http://arXiv.org/abs/1807.07542}{{\tt 1807.07542}}].

\bibitem{Skvortsov:2018uru}
E.~Skvortsov, {\it {Light-Front Bootstrap for Chern-Simons Matter Theories}},
  {\em JHEP} {\bf 06} (2019) 058 [\href{http://arXiv.org/abs/1811.12333}{{\tt
  1811.12333}}].

\bibitem{Goddard:1973qh}
P.~Goddard, J.~Goldstone, C.~Rebbi and C.~B. Thorn, {\it {Quantum dynamics of a
  massless relativistic string}},  {\em Nucl. Phys.} {\bf B56} (1973) 109--135.

\bibitem{Klebanov:2002ja}
I.~R. Klebanov and A.~M. Polyakov, {\it {AdS dual of the critical O(N) vector
  model}},  {\em Phys. Lett.} {\bf B550} (2002) 213--219
  [\href{http://arXiv.org/abs/hep-th/0210114}{{\tt hep-th/0210114}}].

\bibitem{Giombi:2011kc}
S.~Giombi, S.~Minwalla, S.~Prakash, S.~P. Trivedi, S.~R. Wadia and X.~Yin, {\it
  {Chern-Simons Theory with Vector Fermion Matter}},  {\em Eur. Phys. J.} {\bf
  C72} (2012) 2112 [\href{http://arXiv.org/abs/1110.4386}{{\tt 1110.4386}}].

\bibitem{Chalmers:1996rq}
G.~Chalmers and W.~Siegel, {\it {The Selfdual sector of QCD amplitudes}},  {\em
  Phys. Rev.} {\bf D54} (1996) 7628--7633
  [\href{http://arXiv.org/abs/hep-th/9606061}{{\tt hep-th/9606061}}].

\bibitem{Ponomarev:2017nrr}
D.~Ponomarev, {\it {Chiral Higher Spin Theories and Self-Duality}},  {\em JHEP}
  {\bf 12} (2017) 141 [\href{http://arXiv.org/abs/1710.00270}{{\tt
  1710.00270}}].

\bibitem{Siegel:1992wd}
W.~Siegel, {\it {Selfdual N=8 supergravity as closed N=2 (N=4) strings}},  {\em
  Phys. Rev.} {\bf D47} (1993) 2504--2511
  [\href{http://arXiv.org/abs/hep-th/9207043}{{\tt hep-th/9207043}}].

\bibitem{Krasnov:2016emc}
K.~Krasnov, {\it {Self-Dual Gravity}},  {\em Class. Quant. Grav.} {\bf 34}
  (2017), no.~9 095001 [\href{http://arXiv.org/abs/1610.01457}{{\tt
  1610.01457}}].

\bibitem{Bern:1993sx}
Z.~Bern, L.~J. Dixon and D.~A. Kosower, {\it {New QCD results from string
  theory}},  in {\em {International Conference on Strings 93 Berkeley,
  California, May 24-29, 1993}}, pp.~0190--204, 1993.
\newblock \href{http://arXiv.org/abs/hep-th/9311026}{{\tt hep-th/9311026}}.

\bibitem{Bern:1993qk}
Z.~Bern, G.~Chalmers, L.~J. Dixon and D.~A. Kosower, {\it {One loop N gluon
  amplitudes with maximal helicity violation via collinear limits}},  {\em
  Phys. Rev. Lett.} {\bf 72} (1994) 2134--2137
  [\href{http://arXiv.org/abs/hep-ph/9312333}{{\tt hep-ph/9312333}}].

\bibitem{Mahlon:1993fe}
G.~Mahlon, {\it {One loop multi - photon helicity amplitudes}},  {\em Phys.
  Rev.} {\bf D49} (1994) 2197--2210
  [\href{http://arXiv.org/abs/hep-ph/9311213}{{\tt hep-ph/9311213}}].

\bibitem{Mandelstam:1982cb}
S.~Mandelstam, {\it {Light Cone Superspace and the Ultraviolet Finiteness of
  the N=4 Model}},  {\em Nucl. Phys.} {\bf B213} (1983) 149--168.

\bibitem{Brink:1982wv}
L.~Brink, O.~Lindgren and B.~E.~W. Nilsson, {\it {The Ultraviolet Finiteness of
  the N=4 Yang-Mills Theory}},  {\em Phys. Lett.} {\bf 123B} (1983) 323--328.

\bibitem{Beccaria:2015vaa}
M.~Beccaria and A.~A. Tseytlin, {\it {On higher spin partition functions}},
  {\em J. Phys.} {\bf A48} (2015), no.~27 275401
  [\href{http://arXiv.org/abs/1503.08143}{{\tt 1503.08143}}].

\bibitem{Maldacena:2012sf}
J.~Maldacena and A.~Zhiboedov, {\it {Constraining conformal field theories with
  a slightly broken higher spin symmetry}},  {\em Class. Quant. Grav.} {\bf 30}
  (2013) 104003 [\href{http://arXiv.org/abs/1204.3882}{{\tt 1204.3882}}].

\bibitem{Aharony:2012nh}
O.~Aharony, G.~Gur-Ari and R.~Yacoby, {\it {Correlation Functions of Large N
  Chern-Simons-Matter Theories and Bosonization in Three Dimensions}},  {\em
  JHEP} {\bf 12} (2012) 028 [\href{http://arXiv.org/abs/1207.4593}{{\tt
  1207.4593}}].

\bibitem{Aharony:2015mjs}
O.~Aharony, {\it {Baryons, monopoles and dualities in Chern-Simons-matter
  theories}},  {\em JHEP} {\bf 02} (2016) 093
  [\href{http://arXiv.org/abs/1512.00161}{{\tt 1512.00161}}].

\bibitem{Karch:2016sxi}
A.~Karch and D.~Tong, {\it {Particle-Vortex Duality from 3d Bosonization}},
  {\em Phys. Rev.} {\bf X6} (2016), no.~3 031043
  [\href{http://arXiv.org/abs/1606.01893}{{\tt 1606.01893}}].

\bibitem{Seiberg:2016gmd}
N.~Seiberg, T.~Senthil, C.~Wang and E.~Witten, {\it {A Duality Web in 2+1
  Dimensions and Condensed Matter Physics}},  {\em Annals Phys.} {\bf 374}
  (2016) 395--433 [\href{http://arXiv.org/abs/1606.01989}{{\tt 1606.01989}}].

\bibitem{Ponomarev:2016jqk}
D.~Ponomarev and A.~A. Tseytlin, {\it {On quantum corrections in higher-spin
  theory in flat space}},  {\em JHEP} {\bf 05} (2016) 184
  [\href{http://arXiv.org/abs/1603.06273}{{\tt 1603.06273}}].

\bibitem{Metsaev:2005ar}
R.~R. Metsaev, {\it {Cubic interaction vertices of massive and massless higher
  spin fields}},  {\em Nucl. Phys.} {\bf B759} (2006) 147--201
  [\href{http://arXiv.org/abs/hep-th/0512342}{{\tt hep-th/0512342}}].

\bibitem{Chalmers:1998jb}
G.~Chalmers and W.~Siegel, {\it {Simplifying algebra in Feynman graphs. Part 2.
  Spinor helicity from the space-cone}},  {\em Phys. Rev.} {\bf D59} (1999)
  045013 [\href{http://arXiv.org/abs/hep-ph/9801220}{{\tt hep-ph/9801220}}].

\bibitem{Chakrabarti:2005ny}
D.~Chakrabarti, J.~Qiu and C.~B. Thorn, {\it {Scattering of glue by glue on the
  light-cone worldsheet. I. Helicity non-conserving amplitudes}},  {\em Phys.
  Rev.} {\bf D72} (2005) 065022
  [\href{http://arXiv.org/abs/hep-th/0507280}{{\tt hep-th/0507280}}].

\bibitem{Chakrabarti:2006mb}
D.~Chakrabarti, J.~Qiu and C.~B. Thorn, {\it {Scattering of glue by glue on the
  light-cone worldsheet. II. Helicity conserving amplitudes}},  {\em Phys.
  Rev.} {\bf D74} (2006) 045018
  [\href{http://arXiv.org/abs/hep-th/0602026}{{\tt hep-th/0602026}}]. [Erratum:
  Phys. Rev.D76,089901(2007)].

\bibitem{Ananth:2012un}
S.~Ananth, {\it {Spinor helicity structures in higher spin theories}},  {\em
  JHEP} {\bf 11} (2012) 089 [\href{http://arXiv.org/abs/1209.4960}{{\tt
  1209.4960}}].

\bibitem{Bengtsson:2016jfk}
A.~K.~H. Bengtsson, {\it {Notes on Cubic and Quartic Light-Front Kinematics}},
  \href{http://arXiv.org/abs/1604.01974}{{\tt 1604.01974}}.

\bibitem{Ponomarev:2016cwi}
D.~Ponomarev, {\it {Off-Shell Spinor-Helicity Amplitudes from Light-Cone
  Deformation Procedure}},  {\em JHEP} {\bf 12} (2016) 117
  [\href{http://arXiv.org/abs/1611.00361}{{\tt 1611.00361}}].

\bibitem{Benincasa:2007xk}
P.~Benincasa and F.~Cachazo, {\it {Consistency Conditions on the S-Matrix of
  Massless Particles}},  \href{http://arXiv.org/abs/0705.4305}{{\tt
  0705.4305}}.

\bibitem{Benincasa:2011pg}
P.~Benincasa and E.~Conde, {\it {Exploring the S-Matrix of Massless
  Particles}},  {\em Phys. Rev.} {\bf D86} (2012) 025007
  [\href{http://arXiv.org/abs/1108.3078}{{\tt 1108.3078}}].

\bibitem{Konstein:1989ij}
S.~E. Konstein and M.~A. Vasiliev, {\it Extended higher spin superalgebras and
  their massless representations},  {\em Nucl. Phys.} {\bf B331} (1990)
  475--499.

\bibitem{Goroff:1985th}
M.~H. Goroff and A.~Sagnotti, {\it {The Ultraviolet Behavior of Einstein
  Gravity}},  {\em Nucl. Phys.} {\bf B266} (1986) 709--736.

\bibitem{Thorn:2004ie}
C.~B. Thorn, {\it {Renormalization of quantum fields on the lightcone
  worldsheet. 1. Scalar fields}},  {\em Nucl. Phys.} {\bf B699} (2004) 427--452
  [\href{http://arXiv.org/abs/hep-th/0405018}{{\tt hep-th/0405018}}].

\bibitem{Sezgin:2002rt}
E.~Sezgin and P.~Sundell, {\it {Massless higher spins and holography}},  {\em
  Nucl.Phys.} {\bf B644} (2002) 303--370
  [\href{http://arXiv.org/abs/hep-th/0205131}{{\tt hep-th/0205131}}].

\bibitem{Leigh:2003gk}
R.~G. Leigh and A.~C. Petkou, {\it {Holography of the N=1 higher spin theory on
  AdS(4)}},  {\em JHEP} {\bf 0306} (2003) 011
  [\href{http://arXiv.org/abs/hep-th/0304217}{{\tt hep-th/0304217}}].

\bibitem{Colombo:2012jx}
N.~Colombo and P.~Sundell, {\it {Higher Spin Gravity Amplitudes From Zero-form
  Charges}},  \href{http://arXiv.org/abs/1208.3880}{{\tt 1208.3880}}.

\bibitem{Didenko:2013bj}
V.~E. Didenko, J.~Mei and E.~D. Skvortsov, {\it {Exact higher-spin symmetry in
  CFT: free fermion correlators from Vasiliev Theory}},  {\em Phys. Rev.} {\bf
  D88} (2013) 046011 [\href{http://arXiv.org/abs/1301.4166}{{\tt 1301.4166}}].

\bibitem{Didenko:2012tv}
V.~Didenko and E.~Skvortsov, {\it {Exact higher-spin symmetry in CFT: all
  correlators in unbroken Vasiliev theory}},  {\em JHEP} {\bf 1304} (2013) 158
  [\href{http://arXiv.org/abs/1210.7963}{{\tt 1210.7963}}].

\bibitem{Bonezzi:2017vha}
R.~Bonezzi, N.~Boulanger, D.~De~Filippi and P.~Sundell, {\it {Noncommutative
  Wilson lines in higher-spin theory and correlation functions of conserved
  currents for free conformal fields}},  {\em J. Phys.} {\bf A50} (2017),
  no.~47 475401 [\href{http://arXiv.org/abs/1705.03928}{{\tt 1705.03928}}].

\bibitem{Fradkin:1982kf}
E.~S. Fradkin and A.~A. Tseytlin, {\it {Quantum Properties of Higher
  Dimensional and Dimensionally Reduced Supersymmetric Theories}},  {\em Nucl.
  Phys.} {\bf B227} (1983) 252.

\bibitem{Bossard:2015foa}
G.~Bossard and A.~Kleinschmidt, {\it {Loops in exceptional field theory}},
  {\em JHEP} {\bf 01} (2016) 164 [\href{http://arXiv.org/abs/1510.07859}{{\tt
  1510.07859}}].

\bibitem{Joung:2015eny}
E.~Joung, S.~Nakach and A.~A. Tseytlin, {\it {Scalar scattering via conformal
  higher spin exchange}},  {\em JHEP} {\bf 02} (2016) 125
  [\href{http://arXiv.org/abs/1512.08896}{{\tt 1512.08896}}].

\bibitem{Gopakumar:2011qs}
R.~Gopakumar, R.~K. Gupta and S.~Lal, {\it {The Heat Kernel on $AdS$}},  {\em
  JHEP} {\bf 11} (2011) 010 [\href{http://arXiv.org/abs/1103.3627}{{\tt
  1103.3627}}].

\bibitem{Tseytlin:2013jya}
A.~A. Tseytlin, {\it {On partition function and Weyl anomaly of conformal
  higher spin fields}},  {\em Nucl. Phys.} {\bf B877} (2013) 598--631
  [\href{http://arXiv.org/abs/1309.0785}{{\tt 1309.0785}}].

\bibitem{Giombi:2013fka}
S.~Giombi and I.~R. Klebanov, {\it {One Loop Tests of Higher Spin AdS/CFT}},
  {\em JHEP} {\bf 12} (2013) 068 [\href{http://arXiv.org/abs/1308.2337}{{\tt
  1308.2337}}].

\bibitem{Giombi:2014yra}
S.~Giombi, I.~R. Klebanov and A.~A. Tseytlin, {\it {Partition Functions and
  Casimir Energies in Higher Spin $AdS_{d+1}/CFT_d$}},  {\em Phys. Rev.} {\bf
  D90} (2014), no.~2 024048 [\href{http://arXiv.org/abs/1402.5396}{{\tt
  1402.5396}}].

\bibitem{Beccaria:2014jxa}
M.~Beccaria, X.~Bekaert and A.~A. Tseytlin, {\it {Partition function of free
  conformal higher spin theory}},  {\em JHEP} {\bf 08} (2014) 113
  [\href{http://arXiv.org/abs/1406.3542}{{\tt 1406.3542}}].

\bibitem{Beccaria:2014xda}
M.~Beccaria and A.~A. Tseytlin, {\it {Higher spins in AdS$_{5}$ at one loop:
  vacuum energy, boundary conformal anomalies and AdS/CFT}},  {\em JHEP} {\bf
  11} (2014) 114 [\href{http://arXiv.org/abs/1410.3273}{{\tt 1410.3273}}].

\bibitem{Gunaydin:2016amv}
M.~G{\"u}naydin, E.~D. Skvortsov and T.~Tran, {\it {Exceptional $F(4)$
  higher-spin theory in AdS$_{6}$ at one-loop and other tests of duality}},
  {\em JHEP} {\bf 11} (2016) 168 [\href{http://arXiv.org/abs/1608.07582}{{\tt
  1608.07582}}].

\bibitem{Bae:2016rgm}
J.-B. Bae, E.~Joung and S.~Lal, {\it {One-loop test of free SU(N) adjoint model
  holography}},  {\em JHEP} {\bf 04} (2016) 061
  [\href{http://arXiv.org/abs/1603.05387}{{\tt 1603.05387}}].

\bibitem{Skvortsov:2017ldz}
E.~D. Skvortsov and T.~Tran, {\it {AdS/CFT in Fractional Dimension and Higher
  Spin Gravity at One Loop}},  {\em Universe} {\bf 3} (2017), no.~3 61
  [\href{http://arXiv.org/abs/1707.00758}{{\tt 1707.00758}}].

\bibitem{Ponomarev:2019ltz}
D.~Ponomarev, E.~Sezgin and E.~Skvortsov, {\it {On one loop corrections in
  higher spin gravity}},  {\em JHEP} {\bf 11} (2019) 138
  [\href{http://arXiv.org/abs/1904.01042}{{\tt 1904.01042}}].

\bibitem{Ponomarev:2019ofr}
D.~Ponomarev, {\it {From bulk loops to boundary large-N expansion}},  {\em
  JHEP} {\bf 01} (2020) 154 [\href{http://arXiv.org/abs/1908.03974}{{\tt
  1908.03974}}].

\end{thebibliography}
\end{document}